# New attempts to understand nanodiamond stardust

*Ulrich Ott*[A,B], *Astrid Besmehn*[A,C], *Khalil Farouqi*[A], *Oliver Hallmann*[A], *Peter Hoppe*[A], *Karl-Ludwig Kratz*[A], *Karl Melber*[D] *& Anton Wallner*[D,E]

[A]Max Planck Institute for Chemistry, Hahn-Meitner-Weg 1, D-55128 Mainz, Germany;

[B]University of Western Hungary, Savaria Campus, Károlyi Gáspár tér 4, H-9700 Szombathely, Hungary;

[C]Present address: Forschungszentrum Jülich GmbH, Central Division of Analytical Chemistry, D-52425 Jülich, Germany;

[D]University of Vienna, Faculty of Physics, VERA Laboratory, Währinger Strasse 17, A-1090 Vienna, Austria;

[E]Department of Nuclear Physics, Research School of Physics and Engineering, The Australian National University, Canberra, ACT 0200, Australia.

E-mail: uli.ott@mpic.de

Abstract. We report on a concerted effort aimed at understanding the origin and history of the pre-solar nanodiamonds in meteorites including the astrophysical sources of the observed isotopic abundance signatures. This includes measurement of light elements by secondary ion mass spectrometry (SIMS), analysis of additional heavy trace elements by accelerator mass spectrometry (AMS) and dynamic calculations of r-process nucleosynthesis with updated nuclear properties. Results obtained indicate: a) there is no evidence for the former presence of now extinct $^{26}$Al and $^{44}$Ti in our diamond samples other than what can be attributed to silicon carbide and other "impurities"; this does not offer support for a supernova (SN) origin but neither does it negate it; b) analysis by AMS of platinum in "bulk diamond" yields an overabundance of r-only $^{198}$Pt that at face value seems more consistent with the neutron burst than with the separation model for the origin of heavy trace elements in the diamonds, although this conclusion is not firm given analytical uncertainties; c) if the Xe-H pattern was established by an unadulterated r-process, it must have been a strong variant of the main r-process, which possibly could also account for the new observations in platinum.

**Keywords:** stardust – circumstellar matter – isotopic ratios – stars: supernovae type II – ISM: kinematic and dynamics – nanodiamonds

# 1 Introduction

Primitive meteorites contain microscopic grains of stardust, which survived from times before the Solar System was born (Clayton & Nittler 2004; Lodders & Amari 2005; Zinner 2007). Studying these grains, which originated from a variety of stellar sources, is crucial to our understanding of the formation of elements in stars, dust destruction processes in the interstellar medium (ISM), and processes during formation of our Solar System.

Understanding stardust nanodiamonds has not progressed as much as understanding other types of stardust, e.g. silicon carbide and oxide grains. A problem is the small size (average ~2.6 nm), which except for maybe carbon (Heck et al. 2011) does not permit useful single grain isotopic analysis. "Bulk" samples (i.e. millions of grains) yield $^{12}C/^{13}C$ ratios within the range of solar system materials, and also nitrogen isotopes resemble the solar $^{14}N/^{15}N$ ratio as newly re-defined by analyses of solar wind implanted in target materials of the Genesis mission (Marty et al. 2011). Hence, although nominally more abundant than any other known type of pre-solar mineral (except maybe the silicates; e.g., Hoppe 2008), we cannot be certain what abundance fraction of the diamonds is true stardust.

Diagnostic isotopic features are present in several trace elements and suggest a connection to supernovae. This includes xenon-HL (enhancements in p- and r- isotopes), krypton-H (heavy isotopes enhanced) and tellurium (enhancement of the r-only isotopes). Processes suggested to account for the r-isotope (H) enhancements include a neutron burst (Clayton 1989; Meyer et al. 2000), but also a "regular" r-process augmented by an "early" separation between stable end products and radioactive precursors (Ott 1996). While the latter provides a formally better match to the observed Xe and Te isotopic patterns and in principle is also applicable to Kr and to Xe-L, it lacks a credible setting for the separation process to occur.

We have initiated a concerted effort, where we are looking – using secondary ion mass spectrometry (SIMS) - for isotopic features that may be diagnostic for type II supernovae (SN) in light to medium-heavy elements, and – using accelerator mass spectrometry (AMS) – for diagnostic isotopic features in additional heavy trace elements. We also explore whether variants of the high entropy wind (HEW) scenario for the standard r-process may result in the observed unusual enhancements of r-isotopes. Here we report results, which concern the extinct radioactivities $^{26}$Al and $^{44}$Ti (SIMS) and the stable platinum isotopes (AMS). New r-process calculations for xenon are presented that use updated nuclear physics. Partial or preliminary results of this concerted effort have been communicated in Besmehn et al. (2011), Wallner et al. (2011) and Ott et al. (2009, 2010).

## 2  Experimental

Detailed descriptions can be found in dedicated publications (Besmehn et al. 2011; Wallner et al. 2011; Farouqi et al. 2010). Here we give only a short summary of essentials for this combined study.

### 2.1  Nanodiamond samples

Nanodiamond samples were extracted from the meteorites Allende (CV3) and Murchison (CM2) using a combination of chemical and physical separation methods. Three diamond samples were analyzed for radioactivities (Sec. 3.1): a sample from Murchison ("Murch-Std") using the "classical" separation method developed by Amari et al. (1994), another Murchison sample ("Murch-MW") using the microwave technique pioneered by Merchel et al. (2003), and an Allende sample ("Allende") using a variant (high pressure bombs) of the Amari et al. (1994) procedure as described by Braatz et al. (2000). Nanodiamond samples for the AMS measurements (Sec. 3.2) were the Allende AKL, ACL and AMW samples of Merchel et al. (2003). AKL and ACL were prepared by the classical method, with additional purification steps in case of ACL, while AMW was prepared by the microwave method.

## 2.2 SIMS measurements

SIMS measurements were performed with the Cameca IMS 3f instrument at Max Planck Institute for Chemistry. We measured the isotopic composition of Mg together with $^{27}$Al in the search for enhanced $^{26}$Mg from decay of radioactive $^{26}$Al ($T_{1/2}$ = 0.7 Ma). Measurement was in the positive secondary mode using a ~10 nA primary $^{16}$O beam focused into a 10-20 µm size spot. Isobaric interferences were resolved using a mass resolving power M/ΔM ~3500. Except for a slightly lower resolving power (M/ΔM ~3000) the same conditions were used for measurement of Ca (isotopes $^{40}$Ca, $^{42}$Ca, $^{44}$Ca) and $^{48}$Ti in the search for traces of extinct $^{44}$Ti ($T_{1/2}$ = 60 a). $^{48}$Ti was corrected for contributions from $^{48}$Ca assuming a terrestrial (normal) $^{48}$Ca/$^{40}$Ca ratio. For details see Besmehn et al. (2011).

## 2.2 AMS measurements.

Platinum isotopes by AMS were performed at VERA, a 3-MV tandem accelerator at the University of Vienna, which has been designed for AMS measurements up to the heaviest nuclides (Steier et al. 2005; Wallner 2010). Because of a high Pt background the original ion source SNICS (Kutschera et al. 1997) was replaced for our measurements by a new source SN2 of identical design (Priller et al. 2010). Since Pt has a high electron affinity of 2.13 eV, it readily forms negative ions. The terminal voltage was set between 2.8 and 3 MeV. Unlike in "normal" applications of AMS, where one often deals with abundance ratios on the order of $10^{-12}$ or less, in the case of stable Pt isotope measurements there is no need for extreme isobar suppression. Molecular interferences are destroyed during charge exchange, and the only true isobars of the major Pt isotopes (194, 195, 196, 198) from Hg at masses 196 and 198 are not a concern, because Hg does not form negative ions. We did not measure the rare isotopes $^{190}$Pt (normal abundance 0.01 %) and $^{192}$Pt (normal: 0.8 %). The $4^+$ state (yield ~ 6%) was selected after charge exchange for detection. A detailed description of the system and more technical details about our measurements can be found in Melber (2011), Wallner et al. (2011) and references therein.

**2.3  r-process calculations**

Since core-collapse supernovae are the favored source of anomalous trace elements, we studied r-process nucleosynthesis of Xe isotopes in the framework of nucleosynthesis in the high-entropy wind (HEW) of core-collapse supernovae as described previously (Freiburghaus et al 1999; Kratz et al. 2008; Farouqi et al. 2009, 2010). While the traditional aim of these calculations has been reproduction of solar-system r-process abundances, preliminary work (Ott et al. 2009) had indicated that compositions close to the ones observed in the nanodiamonds may be achievable for neutron-rich HEW ejecta under certain conditions of electron fraction ($Y_e$) and entropy (S). Using updated nuclear physics, we investigated in particular the influence of varying the range of entropies on the final results.

# 3  Results and Discussion

**3.1  Extinct radioisotopes**

The results on this subject are presented and discussed in detail in Besmehn et al. (2011) and only a summary is given here. Magnesium and calcium are among the elements that are diagnostic for the stellar sources of presolar grains. Large excesses in $^{26}$Mg can be attributed to the in situ decay of radioactive and now extinct $^{26}$Al ($T_{1/2}$ = 0.7 Ma) which is produced by nuclear reactions in the interior of stars. Aluminium-26 is most abundant in SN grains (SiC X grains, Si$_3$N$_4$, graphite) with inferred initial $^{26}$Al/$^{27}$Al ratios of ~ 10$^{-2}$ to ~ 1 (e.g., Amari et al. 1992; Nittler et al. 1995). Presolar grains from asymptotic giant branch (AGB) stars (SiC mainstream, oxides) have lower $^{26}Al/^{27}Al$ ratios of $\sim 10^{-4}$ to $\sim 10^{-3}$ in SiC (e.g., Hoppe et al. 1994) and $\sim 10^{-4}$ to $\sim 10^{-1}$ in oxides (Nittler et al. 1997, 2008). Another diagnostic isotope is radioactive $^{44}Ti$ ($T_{1/2}$ = 60 years), with supernovae the only stellar sources that produce $^{44}Ti$ (Woosley et al. 1973). Radiogenic $^{44}Ca$ has been found in some SiC X and graphite grains (e.g., Nittler et al. 1996; Besmehn and Hoppe 2003) for which initial $^{44}Ti/^{48}Ti$ ratios between $\sim 10^{-3}$ and $\sim 1$ have been inferred.

The isotope and elemental abundance data of our three diamond samples are given in Table 1, and $^{26}Al$ data are compared to presolar SiC from AGB stars and supernovae in Fig. 1.

Table 1. Mg and Ca isotopic compositions and inferred $^{26}Al/^{27}Al$ and $^{44}Ti/^{48}Ti$ ratios of meteoritic diamond samples.

| Sample | $\delta^{25}Mg^1$ (‰) | $\Delta^{26}Mg^2$ (‰) | $^{26}Al/^{27}Al$ ($10^{-3}$) | $^{26}Al$ (ppb) | $\delta^{42}Ca^1$ (‰) | $\Delta^{44}Ca^2$ (‰) | $^{44}Ti/^{48}Ti$ ($10^{-3}$) | $^{44}Ti$ (ppb) |
|---|---|---|---|---|---|---|---|---|
| Allende | $-13 \pm 7$ | $30 \pm 17$ | $< 1.4$ (2σ) | $< 1.9$ (2σ) | $-12 \pm 5$ | $7 \pm 11$ | $< 1.0$ (2σ) | $< 3.9$ (2σ) |
| Murch-MW | $-18 \pm 7$ | $49 \pm 15$ | $1.00 \pm 0.32$ | $82 \pm 25$ | | | | |
| Murch-Std | $-19 \pm 22$ | $3892 \pm 518$ | $4.48 \pm 0.60$ | $1260 \pm 163$ | $6 \pm 5$ | $7 \pm 13$ | $< 0.47$ (2σ) | $< 13$ (2σ) |

Errors are 1σ.
$^1$ $\delta^iMg = [(^iMg/^{24}Mg)/(^iMg/^{24}Mg)_\odot - 1] \times 1000$; $\delta^iCa = [(^iCa/^{40}Ca)/(^iCa/^{40}Ca)_\odot - 1] \times 1000$
$^2$ $\Delta^{26}Mg = \delta^{26}Mg - 2 \times \delta^{25}Mg$; $\Delta^{44}Ca = \delta^{44}Ca - 2 \times \delta^{42}Ca$

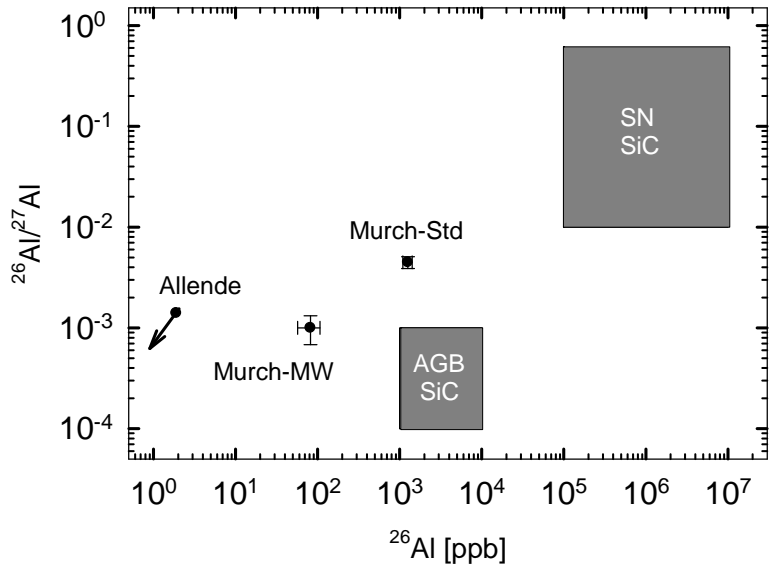

Fig. 1

The $^{25}Mg/^{24}Mg$ ratios of all diamond samples are slightly lower (by about 1-2 %) than solar, most likely due to mass fractionation during analysis. After appropriate correction to the

$^{26}$Mg/$^{24}$Mg ratio, no clear evidence for radiogenic $^{26}$Mg was found for the Allende nanodiamonds, which have δ $^{26}$Mg = 0 within 2σ. However, for the Murch-MW and in particular for the Murch-Std nanodiamonds excesses in $^{26}$Mg of about 5 % and, respectively, of a factor of 4 are seen, indicative of $^{26}$Al decay. The $^{42}$Ca/$^{40}$Ca ratios of the Allende and Murch-Std nanodiamond samples are close to solar. No clear evidence for radiogenic $^{44}$Ca is found, since both diamond samples have δ $^{44}$Ca compatible with zero.

The inferred 2σ upper limits of ~ 10$^{-3}$ for $^{26}$Al/$^{27}$Al and $^{44}$Ti/$^{48}$Ti in the Allende nanodiamonds are lower than those of grains with undisputed SN origin. Whether high $^{26}$Al/$^{27}$Al and $^{44}$Ti/$^{48}$Ti ratios are general features for condensates from SNe, however, remains to be proven. That it may be generally true at least for carbonaceous grains is supported by SN models which predict high $^{26}$Al/$^{27}$Al ratios of >0.01 in C-rich (He/N, He/C) zones (Rauscher et al. 2002). Titanium-44 would be present if matter from the interior zones (Ni, SiS) were mixed to the diamond condensation site in the ejecta.

The Murchison nanodiamonds, in particular those extracted with the standard procedure, show a clear excess of $^{26}$Mg. It is well known that Murchison nanodiamonds prepared by the standard chemical procedure contain fine-grained presolar SiC (Daulton et al. 1996; Yin et al. 2006). It thus seems reasonable to conclude that the observed $^{26}$Mg enrichment is not hosted by the nanodiamonds but represents $^{26}$Al from small amounts of SiC in the Murchison diamond samples. This effect would not have shown up in the Allende sample because the Allende meteorite contains very little SiC (Huss et al. 1995). The amount of radiogenic $^{26}$Mg is smaller for the microwave treated Murchison samples, in which SiC is destroyed to a large extent (Merchel et al. 2003). Some additional input may have come from refractory oxide grains like corundum (Nittler et al. 1996, 2008).

Low ratios $^{26}$Al/$^{27}$Al and $^{44}$Ti/$^{48}$Ti can in principle also be due to dilution with isotopically normal solar-system material, which may be present in our samples. This would not affect the inferred *abundances* of $^{26}$Al and $^{44}$Ti, though, which for the Allende diamond samples at least are orders of magnitude lower than those encountered in the analysis of *bona fide* supernova SiC and graphite (Fig. 1). However, it is not clear to which extent absolute abundances are diagnostic because nanodiamonds may not be able to accommodate as much Al and Ti as SiC and graphite grains

Overall, our $^{26}$Al/$^{27}$Al and $^{44}$Ti/$^{48}$Ti data give no supporting evidence for a SN origin of a major fraction of meteoritic nanodiamonds. Clearly, development of preparation procedures for diamond separates which exclude contamination or contributions from non-diamond material more efficiently would be an important step in order to obtain more reliable data.

## 3.2 AMS of platinum

Platinum is on the peak of the solar r-process abundance distribution (Arlandini et al. 1999), forms efficiently negative ions, which is favorable for AMS, and due to the inability of Hg to form negative ions, has no isobaric interferences at the major isotopes. It is thus the perfect element to search for effects that may be related to the r-process such as the model of Ott (1996). Also in the competing neutron burst model, the abundance is relatively high, since the isotopic pattern is established by a burst acting on a seed of nuclei only few mass units below.

Our results for the three Allende diamond samples are summarized in Table 2. As in Table 1, isotopic ratios are given as δ-values, i.e. deviations from the results obtained for the terrestrial standard. The listed values are averages of results obtained during three beamtimes (Melber 2011). We also include the compositions expected from the two models previously suggested for the explanation of the Xe-H composition, the n burst model (Meyer et al. 2000) and the rapid separation model (Ott 1996).

Obviously, platinum in Allende ACL is indistinguishable from normal, as are the $^{194}$Pt/$^{195}$Pt and $^{196}$Pt/$^{195}$Pt ratios for Allende AKL and AMW. However, we observe enhanced $^{198}$Pt/$^{195}$Pt in the latter two samples. As for the models, both predict strong enhancement of $^{198}$Pt/$^{195}$Pt, and also enhancements by a factor ~2 in $^{196}$Pt/$^{195}$Pt. Where they differ significantly is $^{194}$Pt/$^{195}$Pt. While the neutron burst has only little effect (~ 3 %) on this ratio, the rapid separation in contrast predicts almost complete absence of $^{194}$Pt. This is because the r-process precursor $^{194}$Os has a half life of ~ 6 a, so that almost no decay has occurred at the time of the putative separation after ~2 h between r-process precursors and stable end products, a time that has been suggested based on $^{134}$Xe/$^{136}$Xe in Xe-H (Ott 1996).

Table 2. Results of AMS measurements of Pt isotopes in Allende nanodiamond samples. The last two lines give the predictions from the neutron burst and the rapid separation model.

| Sample | $\delta^{194}$Pt[1] (‰) | $\delta^{196}$Pt[1] (‰) | $\delta^{198}$Pt[1] (‰) |
|---|---|---|---|
| Allende AKL | -19 ± 20 | +5 ± 20 | 51 ± 30 |
| Allende AMW | -7 ± 20 | -8 ± 20 | 71 ± 30 |
| Allende ACL | +1 ± 20 | -12 ± 20 | 2 ± 25 |
| n burst | -31 | 1134 | 14498 |
| rapid separation | -1000 | 886 | 1265 |

Listed results are averages obtained during three beamtimes at VERA. Errors are 1σ.
[1] $\delta^i$Pt= [($^i$Pt/$^{195}$Pt)/($^i$Pt/$^{195}$Pt)$_\odot$ – 1] x 1000

Figure 2a is a three-isotope plot where the sensitive $^{198}$Pt/$^{195}$Pt and $^{194}$Pt/$^{195}$Pt ratios are plotted vs. each other and where the AMS results are shown in comparison with the trends expected from the two models. Figure 2b is a comparison in $^{198}$Pt/$^{195}$Pt vs. $^{196}$Pt/$^{195}$Pt space. Note also that, because Pt is on the r-process abundance peak, there is relatively little difference between the normal composition and the solar r composition obtained by subtracting the s-process contribution (Arlandini et al. 1999).

While ACL plots close to normal, the data for the AKL and AMW separates fit best a mixture of normal Pt with a few percent admixture of Pt from a neutron burst. However, given the analytical uncertainties in comparison with the size (few %) of the effect, the competing rapid separation scenario cannot be completely ruled out either at this time.

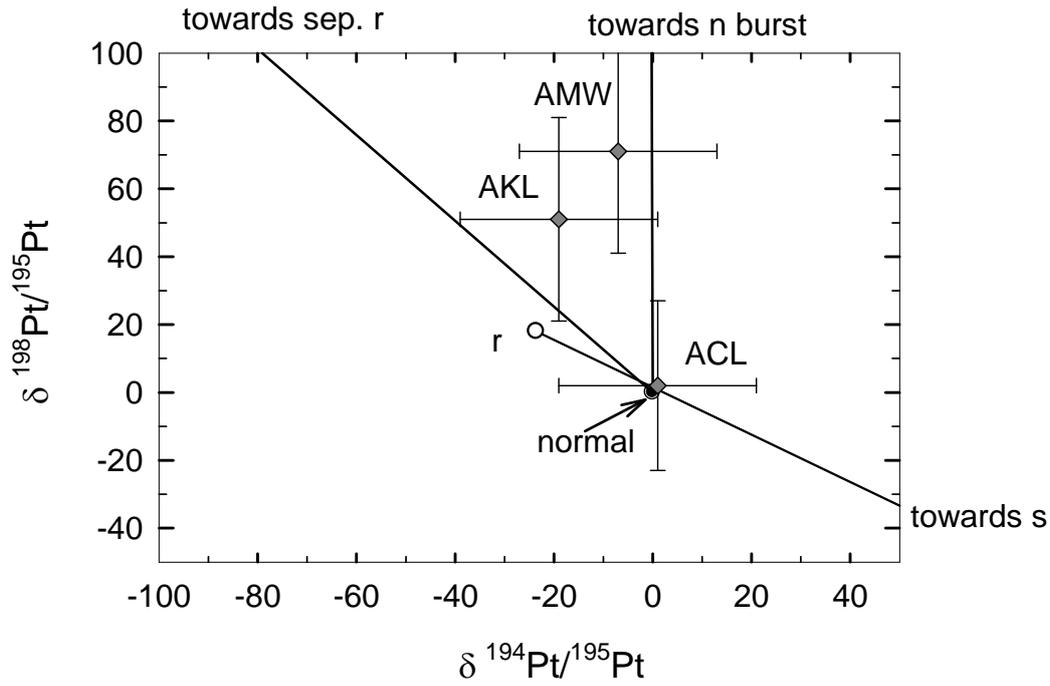

**Fig. 2a**

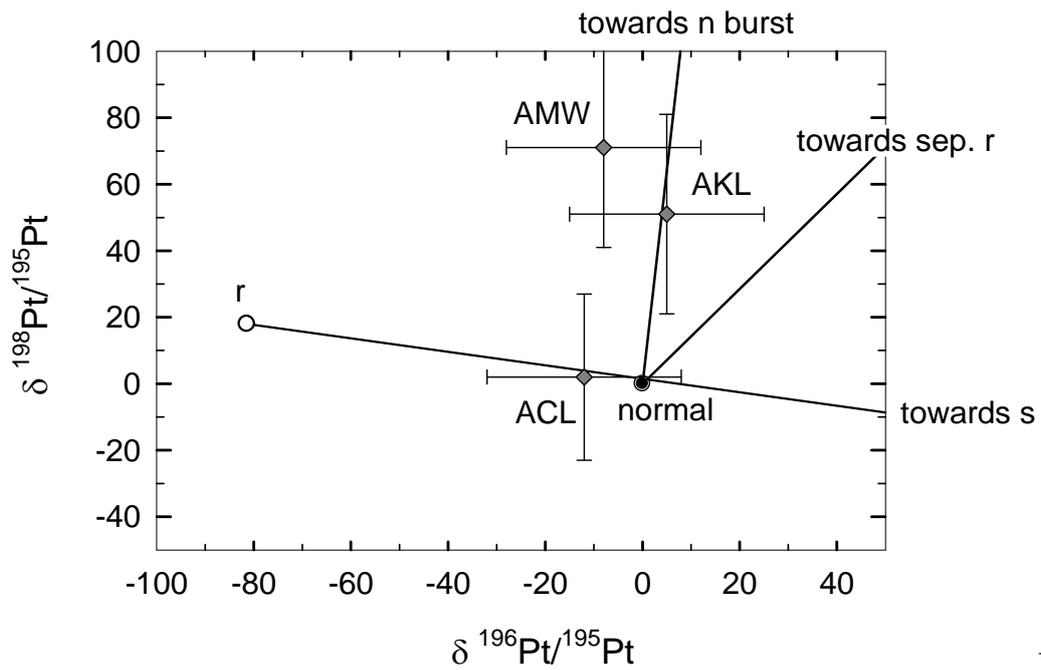

**Fig. 2b**

Another alternative is a "modified" version of the r-process, which we are in the process of further investigating following the recipe of Farouqi et al. (2010). While so far we have concentrated on xenon (discussed in chapter 3.3), preliminary results for moderately neutron rich ejecta ($Y_e = 0.47$) look promising. More detailed study is required, however, to see whether the effects in Xe and Pt can be reproduced at the same time.

That the effect observed in Pt is so much smaller (or even absent in the case of ACL) than in, e.g. Xe is almost certainly due to dilution of Pt carried by the Xe carrier (i.e. nanodiamond) with isotopically normal Pt of solar system origin. Platinum, as the platinum group elements (PGE) in general, is a noble metal and highly resistant to the kind of acid treatments used for extracting the nanodiamonds from the meteorites. In fact, Lewis et al. (1991) found that 12 % of all the iridium in the Allende meteorite survived the acid treatment, ending up in their nanodiamond residue, while Merchel et al. (2003) tried to reduce the PGE content of their diamond samples with only limited success. Nuggets of submicron size consisting of noble metal alloys were discovered in an acid-resistant silicon carbide fraction of the Murchison meteorite by Berg et al. (2009) and it appears likely that similar nuggets (of maybe even smaller size) find their way into nanodiamond separates. While it will be necessary in further work to find a way to suppress the PGE background, we will also expand our efforts to the rare Earth elements in the mass range below the PGE. Background of acid-resistant materials besides diamond should not be such a concern in this case and hence there is a chance that isotopic effects to be observed are larger, but the analysis will be more challenging because of probably lower abundances, lower ionization yield, and isobaric interferences that need to be resolved.

### 3.3  Xe-H and r-process nucleosynthesis

Commonly, r-process nucleosynthesis calculations are performed with the aim of reproducing the solar system r-process pattern (e.g., Farouqi et al. 2010), as obtained from the solar system abundances after subtraction of the s-process contribution, as e.g. by Arlandini et al. (1999). In contrast to detailed isotopic patterns, comparison of *elemental* abundances, as, e.g., in the case of stellar observations of metal-poor r-process-rich stars (e.g. Farouqi et al. 2009), requires less overall accuracy and is less subject to uncertainties in nuclear physics parameters, which to some extent average out. On the other hand, isotopic variations that may be attributed to the s-process are commonly based on comparison with the "observed" (i.e. s-process corrected) r-process patterns (e.g., Schönbächler et al. 2003; Carlson et al. 2007, Hidaka & Yoneda 2011) rather than the results of r-process calculations. Given progress has been made with respect to the nuclear properties of nuclei involved in the creation of the A~130 r-process abundance peak, we have started on an exploratory study of possible modifications of the r-process proper that may result in the observed Xe-H pattern (Ott et al. 2009), investigating the parameter space allowed by the high entropy wind (HEW) scenario (Farouqi et al. 2010). We have pointed out previously, in the context of the observed abundances of extinct radionculides $^{182}$Hf and $^{129}$I in the early Solar System, the sensitivity of nuclide abundance ratios to the range of entropies included in the calculation of the yields, heavier nuclides being primarily produced in the higher entropy range (Ott & Kratz 2008). Similarly, production of the heaviest Xe isotope $^{136}$Xe is most effective at higher entropy than production of the lightest r-process Xe isotope $^{129}$Xe.

This is illustrated in Fig. 3, where we have also included updated nuclear physics in comparison to Farouqi et al. (2010) and the results presented in Ott et al. (2009). Significant changes derive from "new nuclear structure" as inferred from the experiments of Pfeiffer et al. (2001) concern the yields $P_n$ for β-delayed neutron emission. Most important are changes for $^{134}$Cd (new: $P_{1n}$ = 75 %, $P_{2n}$ = 18 %, $P_{3n}$ = 7 % vs. old: $P_{1n}$ = 30 %, $P_{2n}$ = 42 %, $P_{3n}$ = 3 %), $^{133}$Ag (new: $P_{1n}$ = 3.5 %, $P_{2n}$ = 2.2 %, $P_{3n}$ = 93 % vs. old: $P_{1n}$ = 9 %, $P_{2n}$ = 68 %, $P_{3n}$ = 20 %), and $^{132}$Pd (new: $P_{1n}$ = 3.5 %, $P_{2n}$ = 2.6 %, $P_{3n}$ = 90 % vs. old: $P_{1n}$ = 14 %, $P_{2n}$ = 32 %, $P_{3n}$ = 44 %). All results shown here as well as in Figs. 4a,b have been obtained using the EFTSI-Q mass model for the involved neutron-rich unstable nuclei and for "standard assumptions" with regard to the electron fraction ($Y_e$ = 0.45) and expansion velocity ($x_{exp}$ = 7500 km/sec) in the HEW model.

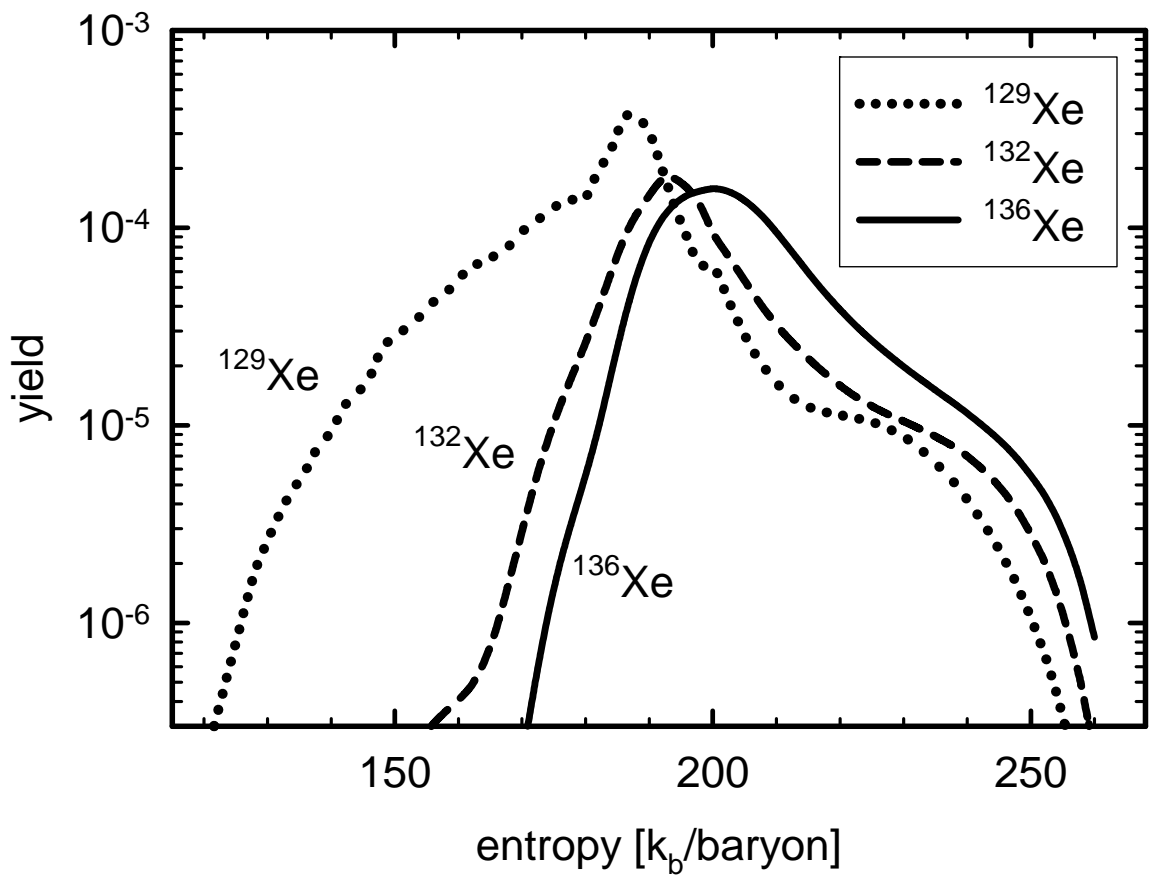

**Fig. 3**

Resulting isotopic ratios for various ranges of entropy are plotted in Fig. 4a, where the resulting patterns are compared with the pattern observed / derived for Xe-H in the meteoritic nanodiamonds (Ott 1996). It is apparent that by including the entropy range below 200 $k_b$/baryon the light isotopes (129, 131, 132) are overproduced relative to $^{136}$Xe and the H-Xe isotopic pattern, while excluding the range below 210 $k_b$/baryon leads to $^{134}$Xe/$^{136}$Xe ratios higher than in Xe-H.

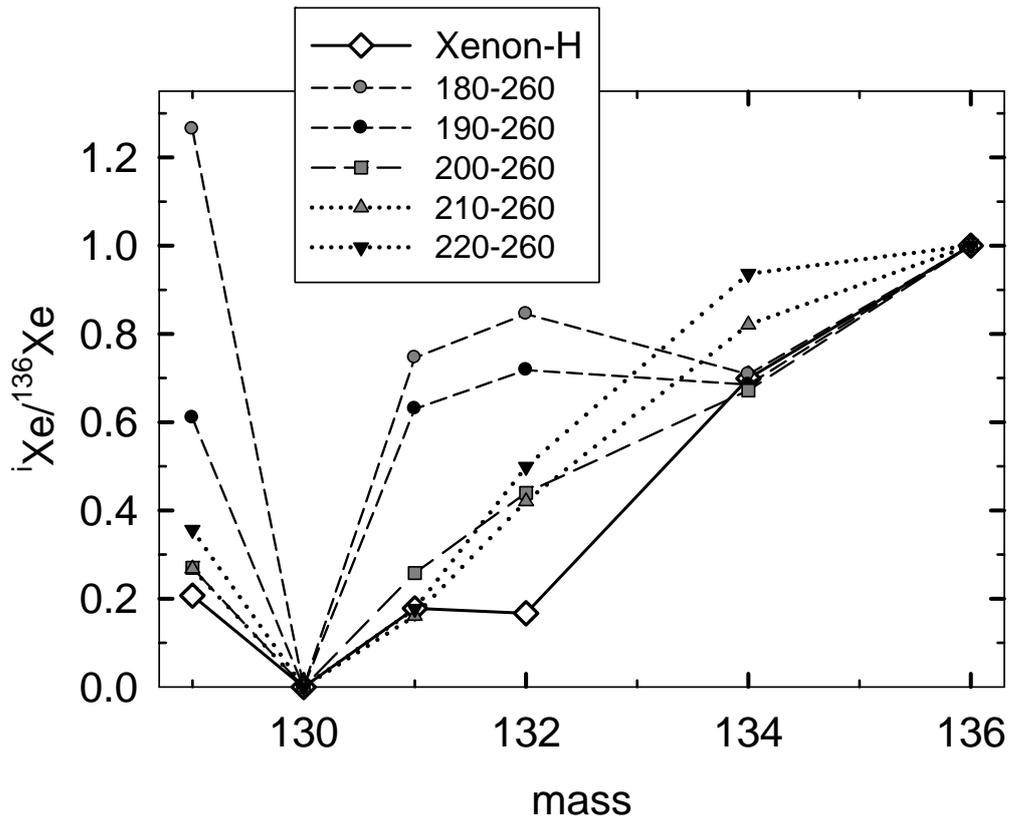

**Fig. 4a**

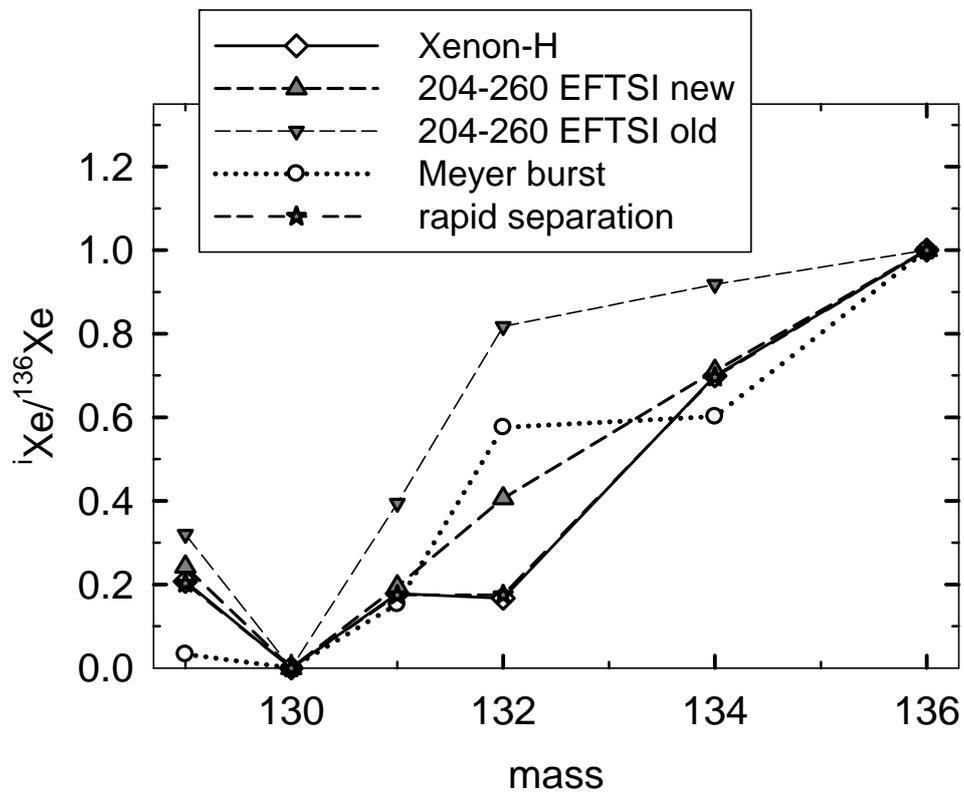

**Fig. 4b**

A kind of "best fit" is the range S = 204-260 $k_b$/baryon. This is shown in Fig. 4b, where the resulting pattern is compared with Xe-H and the predictions of the neutron burst and rapid separation models. Also shown for comparison are the corresponding results obtained in the earlier calculations (Ott et al. 2009) that were using the "old" physics. Obviously, the results of our "restricted entropy range" model, which essentially corresponds to a strong "main" r-process component, are at least as good a match as the neutron burst model of Meyer et al. (2000).

We note, however, that even with our new physics input, significant uncertainties remain in the application of HEW r-process nucleosynthesis to the problem of the origin of Xe-H. In particular, the results sensitively depend on the mass model that is used. The calculations presented here were all obtained with the EFTSI-Q version (Pearson et al. 1996) of the ETFSI (Aboussir et al. 1995) mass model. This model takes into account the experimental indications for N=82 shell-quenching and corrects the local overestimation of the N=82 shell strength below doubly-magic $^{132}$Sn (see, e.g. Dillmann et al. 2003). The FRDM mass model (Möller et al. 1995) as an alternative, e.g., which predicts an increase of the N=82 shell-strength with decreasing atomic number, results in significantly worse agreement. Nevertheless, even with the quenched ETFSI-Q masses, it is not easy to find conditions where the Xe isotopic r-process abundances are produced in solar system proportions (which ideally they should, if the full entropy range is taken, since in this case the elemental abundance is reproduced). In our ongoing explorations of the HEW model we are considering a further local improvement in the nuclear-data input, as well as moderate changes in the astrophysical model parameters of the HEW expansion velocity and stellar temperature (Farouqi & Kratz, in preparation). With respect to nuclear data, new "stellar half-lives" deduced from the known ground-state and isomeric-state ß-decays for the N=82 odd-proton waiting-point isotopes $^{131}$In and $^{129}$Ag are taken into account (see, e.g. Arndt et al. 2005, 2011). Concerning stellar parameters, attempts to slightly shift the A=130 abundance peak by reducing the expansion velocity $v_{exp}$ will mimic a so-called "hot" r-process variant, which will decrease the $^{129}$Xe and increase the $^{136}$Xe abundance (Arndt et al. 2011, Farouqi & Kratz 2011). Not least, it is important to take note that there is also uncertainty in the exact composition of Xe-H. This is discussed in detail in the final chapter (Sec. 3.4).

We also note that there are more alternatives that may be worth exploring. For example, in the Rauscher et al. (2002) model, neutron capture reactions leading to $^{136}$Xe being more abundant than $^{131}$Xe and $^{132}$Xe are also at work in specific zones of the supernova. In our discussion of Ba isotopes in supernova SiC grains (Marhas et al. 2007), we have compared our data with this kind of "neutron burst", which occurs in the C/O zone of the 25 M$_\odot$ model of these authors. When it comes to Xe, it is only in the outer part of that zone that $^{136}$Xe becomes more abundant than $^{132}$Xe and $^{131}$Xe, but this is accompanied by also rather low (compared to Xe-H) $^{134}$Xe/$^{136}$Xe. Nevertheless it is possible to pick a range where the overall pattern is similar to that of the Meyer et al. (2000) neutron burst. Another interesting case is that of r-process nucleosynthesis in neutron star mergers (rather than the explosion of a massive star investigated in our scenario) explored by Goriely et al. (2011). Interestingly the abundance peak is shifted to higher mass relative to r-process solar abundances (see. Fig. 4 in Goriely et al. 2011), but the peak in their calculation is at $^{134}$Xe, not at $^{136}$Xe as in Xe-H. A common problem to all the models discussed here and shown in Fig. 4b (save the rapid separation) is that they produce too much of $^{132}$Xe relative to $^{136}$Xe and the other neutron-rich Xe isotopes. None of them reproduces the dip in abundance when going from $^{131}$Xe to $^{132}$Xe.

## 3.4 The true composition of Xe-H

We have alluded to previously (Sec. 3.3) that uncertainties exist also in the true isotopic composition of Xe-H. It needs to be realized in this context is that in approaches as discussed in Sec. 3.3, the aim is to reproduce a "pure" Xe-H composition that is characterized by zero abundance of s-only $^{130}$Xe. A common way of doing this has been subtracting from the most extreme (highest $^{136}$Xe, lowest $^{130}$Xe) composition a known component thought to contain all of the $^{130}$Xe. Currently available data are characterized by a most extreme measured

$^{136}$Xe/$^{130}$Xe ratio of 4.534 (Ott 2002). Following the standard approach also used in Ott (1996), we subtracted for the discussion in Sec. 3.3 Xe with solar wind composition (Pepin et al. 1995) from this "measured HL", assuming all $^{130}$Xe is part of the solar wind component.

However, this approach is in no way unique, and in particular the assumption of "background" Xe having solar composition may be questioned. Naturally, for the isotopes that have a large background contribution, $^{129}$Xe, $^{131}$Xe and $^{132}$Xe (but not $^{134}$Xe) the abundance in the H component is rather sensitive to this assumption. On the other hand, nanodiamonds contain also the Xe-P3 noble gas component released at low temperature (Huss & Lewis 1994) and both P3 and HL gases most likely were introduced by ion implantation (e.g., Lewis et al. 1989; Verchovsky et al. 2003). Experimental investigations of ion implantation into nanodiamond (Koscheev et al. 2001) indicate that the implanted gases show a bimodal temperature release –as observed in the meteoritic nanodiamonds – which then suggests that the high-temperature $^{130}$Xe-containing component present besides the ($^{130}$Xe-free) pure Xe-H is the same P3 component as observed at low temperature (Huss et al. 2008). In other words, it is more likely that the "background" Xe has the P3 rather than solar composition.

Table 3. Composition of Xe-H derived using different assumptions concerning the composition of background Xe. The last line gives the "source" composition of Xe-H derived by assuming that during implantation not only background P3-Xe was mass fractionated, but Xe-H as well.

| background Xe | $^{129}$Xe/$^{136}$Xe | $^{131}$Xe/$^{136}$Xe | $^{132}$Xe/$^{136}$Xe | $^{134}$Xe/$^{136}$Xe |
|---|---|---|---|---|
| solar | 0.207 | 0.178 | 0.167 | 0.699 |
| P3 | 0.112 | 0.114 | 0.074 | 0.677 |
| fract. P3 | 0.144 | 0.098 | 0.027 | 0.671 |
| fract. P3, Xe-H at source | 0.154 | 0.103 | 0.028 | 0.684 |

To complicate matters further, the Koscheev et al. (2001) results – similar to observations in other ion implantation studies using different materials (e.g., Nichols et al. 1992; Hohenberg et al. 2002) – indicate that the high temperature part of the implanted xenon is mass fractionated relative to the starting composition and that measured in the low temperature release, by about 1 % per a.m.u (Huss et al. 2008), which would then be the appropriate composition to subtract for deriving the pure from the measured Xe-H composition. In addition, if the pure Xe-H was introduced by the same kind of ion implantation, the thus derived Xe-H would differ from that in the source by the same kind of fractionation factor. Resulting compositions after including the various changes are listed in Table 3. Importantly, the already rare isotopes 131 and 132 are further diminished in this process, $^{132}$Xe in particular coming down to almost zero abundance. This makes it even more difficult to find a match in the models.

# 4    Summary

In a concerted effort aimed at understanding the origin and history of the pre-solar nanodiamonds in meteorites we have performed isotopic analysis by SIMS of Mg and Ca to search for the signatures of extinct radioactive $^{26}$Al and $^{44}$Ti. We also measured the isotopic composition of Pt using AMS to search for isotopic effects connected to isotopically anomalous noble gases in order to further improve the understanding of the origin of the Xe-H component. Complementary, we performed dynamic calculations of r-process nucleosynthesis with updated nuclear properties to explore whether variations to the r-process can result in the isotopic pattern Xe-H observed in the diamonds and can thus be an alternative to previously suggested scenarios.

   Our results indicate:

a) There is no evidence for the former presence of now extinct $^{26}$Al and $^{44}$Ti in our diamond samples (other than what can be attributed to silicon carbide and other "impurities") that would uniquely tie them to a supernova origin.

b) Platinum in "bulk diamond" is characterized by an overabundance of r-only $^{198}$Pt that – unlike effects in Xe and Te - seems more consistent with the neutron burst than with the separation model for the origin of heavy trace elements in the diamonds. However, the size of the errors does not allow a firm conclusion. Preliminary results for the high entropy wind r-process scenario suggest this may be also an alternative.

c) If the Xe-H pattern was established by an unadulterated r-process, it must have been a strong "main" r-process variant of the neutron capture process that produces the r-"residuals" in solar system proportions.

Essential for further progress will be the preparation of purer diamond separates. Only most recently it has been recognized that the "diamond" separates contain a substantial fraction of "glassy carbon", which is chemically similarly resistant as diamond but difficult to detect by electron microscopic methods (Stroud et al. 2011) and also (or preferentially?) may be the host to the trace elements. If the level of trace impurities can be reduced, this should enable us to set more stringent limits on the abundance of the radioactivities $^{26}$Al and $^{44}$Ti. Moreover, residues that contain smaller contribution of noble metal nuggets will hopefully result in the detection of larger and thus more diagnostic effects in platinum and/or other platinum group elements. Complementary measurements of Rare Earth elements that are less susceptible to contamination but more challenging for AMS may also contribute to a better understanding the stellar sources of trace elements in the diamond separates. Finally, further exploration of the parameter space in r-process nucleosynthesis coupled with improved knowledge of the properties of key nuclei may provide additional insight. A particular challenge lies also in understanding the process by which the trace elements were introduced into the diamonds and/or glassy carbon, what the isotopic composition of the "background "of these elements is and the kind of isotopic fractionation that may have happened during the introduction.

**Acknowledgments**

We thank Silke Merchel and Christa Sudek for preparation of the nanodiamond residues. Part of this work was funded by the Austrian Science Fund (FWF): project numbers AP20434 and AI00428 (CoDustMas, Eurogenesis).

**Figure Captions**

**Figure 1.** Inferred $^{26}$Al/$^{27}$Al ratios and inferred abundances of extinct $^{26}$Al. Data for the diamond separates are compared to the ranges observed for SiC grains from AGB stars and SiC grains from supernovae.

**Figure 2.** Three isotope plot of $^{198}$Pt/$^{195}$Pt vs. $^{194}$Pt/$^{195}$Pt (a) and $^{198}$Pt/$^{195}$Pt vs. $^{196}$Pt/$^{195}$Pt (b) given as permil deviation from a terrestrial standard. Shown are the AMS data for the three Allende nanodiamond separates, together with the trends for mixing isotopically normal Pt with Pt having the composition predicted by the neutron burst and rapid separation models (Meyer et al. 2000; Ott 1996). Also shown is the composition of solar system r-process Pt and the mixing line between s- and r-process Pt (Arlandini et al. 1995).

**Figure 3.** Calculated yields as function of entropy for r-process production in the HEW scenario of selected Xe isotopes: $^{129}$Xe, $^{132}$Xe and $^{136}$Xe. With increasing mass, production peaks at increasing entropy.

**Figure 4.** Isotopic abundances (normalized to $^{136}$Xe = 1) of r-process xenon produced in the HEW model for selected entropy ranges as indicated in the legend. a) Patterns produced in various "interesting" entropy ranges (in $k_b$/baryon) are compared with each other and with the Xe-H composition. b) The "best range" (204-260 $k_b$/baryon) composition is compared to results of the identical calculation using "old" physics (see text), and to Xe-H and predictions from competing models (Meyer et al. 2000; Ott 1996).